\documentstyle[prl,aps,epsf,float,times]{revtex}
\begin{document}
\draft
\twocolumn[\hsize\textwidth\columnwidth\hsize\csname@twocolumnfalse\endcsname
\title{ESR Modes in CsCuCl$_{3}$ in Pulsed Magnetic Fields}
\author{S. Schmidt, B. Wolf, M. Sieling, S. Zvyagin, I. Kouroudis, B. L\"{u}thi}
\address{Physikalisches Institut, Universit\"{a}t Frankfurt
D-60054 Frankfurt, Germany}
\date{\today} \maketitle
\begin{abstract}
We present ESR results for 35 -- 134GHz in the antiferromagnet CsCuCl$_{3}$ at
$T=1.5 \text{K}$. The external field is applied perpendicular to the hexagonal $c$-axis. With our
pulsed field facility we reach $50 \text{T}$ an unprecedented field for low temperature ESR. We
observe strong resonances up to fields close to the ferromagnetic region of $\approx 30 \text{T}$. These
results are discussed in a model for antiferromagnetic modes in a two-dimensional frustrated
triangular spin system.\\
\end{abstract}
]

CsCuCl$_{3}$ belongs to the hexagonal ABX$_{3}$ type compounds in which linear chains
of face sharing octahedra (BX$_{3}$) are formed along the {\it c}-axis. A strong intrachain
ferromagnetic coupling and smaller interchain antiferromagnetic interaction give rise to a
triangular spin arrangement in the {\it c}-plane below $T_{N}=10.5 \text{K}$. A cooperative
Jahn-Teller transition at 423K distorts the CuCl$_{6}$ octahedra \cite{kroese}, but the structure
remains hexagonal with a tripled {\it c}-axis unit cell. The low symmetry of the local structure
leads to the antisymmetric Dzyaloshinsky - Moriya interaction along the chains and gives a
helical spin structure along the {\it c}-axis with a long period \cite{adachi}.

More recently the research for this compound focussed on two topics: On the one hand an
interpretation of the beautiful high field magnetization experiments \cite{nojiri} using quantum
fluctuations in this frustrated spin system  for $B\parallel c$ \cite{nikuni} and  for $B\perp c$
\cite{jacobs} has been given. On the other hand a detailed investigation of the magnetic phases near
$T_{N}$ \cite{weber} elucidating the chiral XY state has been carried out.

ESR experiments have been performed for both $B\parallel c$ and  for $B\perp c$ in the
frequency range 35 -- 280GHz \cite{palme} and 90 -- 380GHz \cite{ohta} for fields up to 14T in the low
temperature range $T<T_{N}$. In the case $B\parallel c$ one observes the
two branches of the antiferromagnetic resonance modes which can be calculated for
$B<B_{c}$ taking the helical spin structure into account \cite{tanaka}. At $B=B_{c}$
(12.5T at 1.1K) there is a small  jump in the magnetization curve \cite{nojiri} and a shift of the
upper resonance frequency \cite{ohta}. $B_{c}$ is the critical field where the umbrella type spin
structure changes to a coplanar structure \cite{nikuni}.

For $B\perp c$ the physics is more complicated since the magnetic field breaks the
axial symmetry of the system. The helical spin structure for $B=0$ becomes a  distorted
helix for $B>0$. It is possible to get from this incommensurate structure to a
commensurate one at a critical field estimated theoretically $B_{c1}=10 - 11\text{T}$ \cite{jacobs}
(but see below). Inclusion of quantum fluctuations on the commensurate state establishes in
the magnetization a small plateau at $B_{c1}$ as observed experimentally \cite{nojiri}. The low
temperature ESR experiments show again two antiferromagnetic resonance branches with the
lower one tending to zero at $B_{c1}$ \cite{palme}. Apart from strong resonances we observe
also smaller ones in the vicinity of $B_{c1}$ \cite{palme,ohta}. The interpretation of these results is
more complicated than in the $B\parallel c$ case. So far simple models like a spinwave
calculation for the triangular frustrated planar spins \cite{rastelli} or a classical resonance calculation
for this system \cite{palme} have been performed.

We performed ESR experiments in very high magnetic fields up to 50T for the $B\perp c$ configuration
(For the $B\parallel c$ geometry the resonance frequencies are smaller than
35GHz in the high field region and therefore not attainable for us). We used the pulsed field
facility of the Frankfurt High Magnetic Field Laboratory \cite{wolf} to obtain a maximum field of
50T with a typical pulse length of 30ms. We detect ESR lines in the frequency range 35 --
134GHz with a liquid Helium cooled fast InSb hot electron bolometer. The microwave
radiation was generated by Gunn oscillators and Impatt diodes with frequency multipliers. The
radiation is transmitted twice through the specimen by reflecting on a Al-coated ceramic disk
below the specimen and brought to the bolometer with directional couplers. Pulsed field ESR
has been performed before in the pioneering work of Date {\it et al.\/} \cite{date}.

\begin{figure}
\centerline{\epsfxsize=3in\epsffile{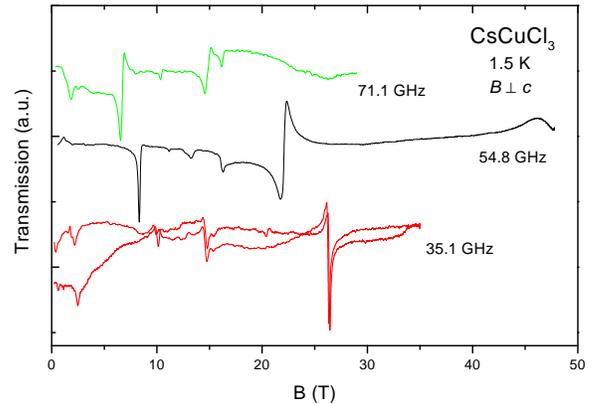}}
\caption{Transmission of the ESR signal as a function of magnetic field B  for three different
frequencies at $T=1.5 \text{K}$. The strong resonances exhibit large dispersive effects.}
\label{fig:1}
\end{figure}

In fig.1 we show typical ESR signals for 35.1, 54.8 and 71.1GHz at 1.5K in the field range
up to 48T. The signal amplitude is plotted for the rising section of the field pulse. The
resulting thermal effects due to eddy currents of the sample holder or due to magnetocaloric
effects in the sample are almost negligible \cite{wolf} as seen from the 35GHz curve where we
show the result for the ascending and descending section of the magnet pulse. From the
temperature dependence of the resonances to be discussed below we expect at most a heating
effect of $<0.5 \text{K}$ for a 50T pulse. For each frequency we observe two large resonances
(e.g. at 8T and 22T for the 54.8GHz result) and up to three smaller resonances (e.g. at
11T, 13T and 16T for the 54.8GHz resonance) which correspond to the small resonances
mentioned above.

In fig.2 we plot the observed resonances as a function of magnetic field. The full symbols
indicate the strong resonances, the crosses the small ones and the open circles resonances
from ref.\cite{palme}. In the region of overlap for the data of ref.\cite{palme} and our new one the agreement is very
good. As to the smaller resonances they partly agree with older results. We believe that these
smaller resonances depend on the actual domain state and sample quality. They can also arise
as excitations from the complicated ground state especially in the region of the
incommensurate - commensurate transition and are very difficult to calculate. In the following
we discuss only the large resonances. In fields up to 14T they have been observed already by
different groups and in different crystals.  A remarkable aspect of fig.2 is the absence of
resonances for fields $B>30\text{T}$ for the frequency range 35 -- 134GHz. This is shown
e.g. in fig.1 for one frequency 54.8GHz.

\begin{figure}
\centerline{\epsfxsize=3.3in\epsffile{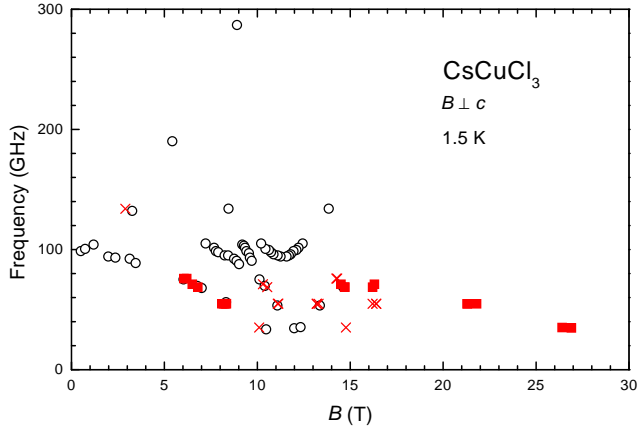}}
\caption{Frequency versus magnetic field $B$ for $B\perp c$ at $T=1.5 \text{K}$.
open circles: results from ref. [7], crosses: small resonances, closed squares: strong resonances
from this work. Note the good agreement with former results in the region of
overlap.}
\label{fig:2}
\end{figure}

The calculation of these resonances for the field configuration $B\perp c$ is a very
difficult problem. The evaluation of the ground state spin arrangement is already complicated
as discussed above. Even in molecular field approximation this is a nontrivial problem, even
more so with the important inclusion of quantum fluctuations in the commensurate and
incommensurate state \cite{jacobs}. For the discussion of our resonance results we therefore use
simplified models.

There are two models which have been discussed and applied to the present state, a spinwave
model \cite{rastelli} and a classical calculation of the antiferromagnetic resonance modes \cite{palme} (i.e. a
$q=0$ spinwave calculation). Both models assume a two-dimensional triangular frustrated
Heisenberg spin model, an approximation which neglects the spiral spin chains along the {\it
c\/}-axis. A strong uniaxial anisotropy field of $B_{a} \approx 0.36 \text{T}$ confines the spins to
the hexagonal plane. Since these models give the salient features of ESR experiments
performed up to $B<14 \text{T}$ \cite{palme} and since the magnetization results \cite{nojiri} give also the
same three regions of spin arrangements discussed in these models ($0<B<B_{c1}$, $B_{c1}<B<3B_{c1}$,
$3B_{c1}<B$ we would like to know how these calculations describe our results at higher fields. In fig.3 we
show the results for the antiferromagnetic resonance model (with $B_{ex }=B_{c1}=10 \text{T}$, $B_{a}=0.36\text{T}$)
together with experimental results from fig.2 including only the strong resonances.
The equations for the antiferromagnetic resonance for the lower mode are:

\[
\begin{array}{lcl}
\omega = \gamma \sqrt{3B_{a}B_{ex}-B^{2}\frac{B_{a}}{B_{ex}}-2B_{a}B}
&\text{for}& B<B_{c1}\\
\\
\omega = \gamma \sqrt{3B_{a}B_{ex}+\frac{B^{2}}{2}-\sqrt{I}}
&\text{for}&B_{c1}<B<B_{c2}\\
\\
\multicolumn{3}{l}
{\omega = \gamma\sqrt{B^{2}+2BB_{a}-6BB_{ex}-6B_{a}B_{ex}+9B_{ex}^{2}}}
\\
\\
&\text{for} &B>B_{c2}\\
\end{array}
\]

where $\gamma$ is the gyromagnetic ratio and

\begin{eqnarray*}
I&=&B^{4}\left(\frac{1}{4}+\frac{B_{a}^{2}}{2B_{ex}^{2}}+\frac{3B_{a}}{4B_{ex}}\right)
    -B^{2}\left(5B_{a}^{2}+\frac{9B_{ex}B_{a}}{2}\right)\\
 & &{}+\frac{27B_{ex}^{2}B_{a}^{2}}{2}
    +\frac{27B_{ex}^{3}B_{a}}{4}
\end{eqnarray*}

\begin{figure}
\centerline{\epsfxsize=3.3in\epsffile{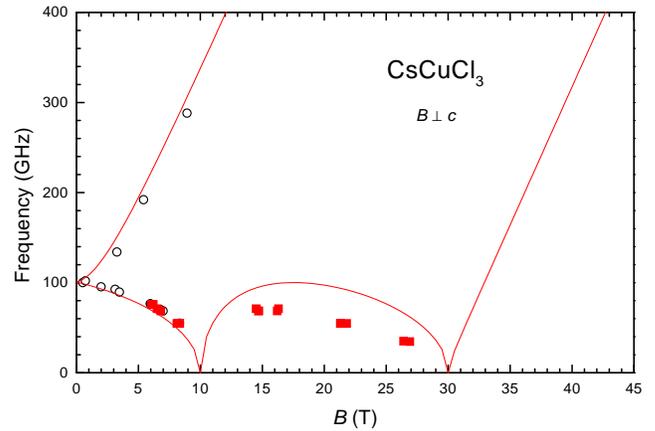}}
\caption{Fit of the strong resonances from fig.2 with the antiferromagnetic resonance model (full
line) as discussed in the text.}
\label{fig:3}
\end{figure}

Whereas the resonances are well described below $B_{c1 }$ by this model as already
shown in ref.\cite{palme} the agreement is only qualitative for $B > B_{c1}=10 \text{T}$.
In addition no resonance has been observed in the saturated region for $B > B_{c2}= 3 B_{c1}= 30 \text{T}$.

For this latter effect one of the explanations could be the following polarisation features of our
ESR experiment. In the high field region we have a ferromagnetic state and the resonance
field has to be a transverse field. In our resonance arrangement we have a Faraday geometry
but with wavelengths of the order of the tube diameter. Therefore we have cavity modes with
the microwave fields $b_{rf}$ parallel and perpendicular to the applied field. Our
crystal was placed in a position with predominantly parallel $b_{rf}$ field. Therefore
the absence of transverse resonances for $B>30 \text{T}$ in this frequency regime follows.

We also investigated the temperature dependence of the resonance frequencies. As an
example we investigated the 35GHz resonance in the temperature region 1.5 -- 3.8K. We
find a strong decrease of the resonance field. On going from 1.5 to 3.8K the field changes
from 26.4 to 24.8T for $B_{c1}<B<B_{c2}$. This follows from
the decrease of the exchange and anisotropy fields with rising temperature.

Recent neutron diffraction experiments \cite{nojiri2} demonstrate a transition from incommensurate to
commensurate spin structure at 17T for 4.2K and not at $B_{c1}$ as expected before.
This result gives support for our model assumption for quasi two-dimensional spin
arrangements for $B>17 \text{T}$. A theory for the excitations including Dzyaloshinsky --
Moriya interaction and interplanar exchange has not been performed yet.

In conclusion we have demonstrated the feasibility of ESR experiments in pulsed fields up to
50T in CsCuCl$_{3}$ at low temperatures. We found resonances in the field region where
the triangular spin arrangement is collapsed into partly collinear spins ($B>B_{c1}$)
and that these resonances show a softening towards higher fields where the spins
ultimately align to a ferromagnetic arrangement. However the resonance frequencies lie
appreciably below our calculated modes for a frustrated two-dimensional triangular lattice.
We hope that these high field ESR data will stimulate further theoretical investigations of this
high field phase and its excitations.

The crystal used in this investigation was grown by Dr. H.A.~Graf, Hahn-Meitner Institute,
Berlin. We acknowledge gratefully helpful discussions with H.~Shiba and H.~Motokawa.
S.Z. acknowledges support of the A.v.Humboldt foundation. This research was supported by
BMBF grant No 13N 6581 and by the DFG through SFB 252.


\vspace*{-4ex}
\begin{references}
\vspace*{-11ex}

\bibitem{kroese}
C.J.~Kroese, J.C.M.~Tindemans, W.J.A.~Masskant, Solid State Comm. {\bf 9}, 1707
(1971); U.~F\"{o}rster, H.A.~Graf, U.~Schotte, U.~Stuhr, J.Phys.C.M. {\bf 9}, 1067
(1997).

\bibitem{adachi}
K.~Adachi, N.~Achiwa, M.~Mekata,  J.Phys.Soc.Jpn. {\bf 49}, 545 (1980).

\bibitem{nojiri}
H.~Nojiri, Y.~Tokunaga, M.~Motokawa, J.Phys. (Paris) {\bf 49,} 1459 (1988) Suppl. C8

\bibitem{nikuni}
T.~Nikuni, H.~Shiba, J.Phys.Soc.Jpn. {\bf 62}, 3268 (1993).

\bibitem{jacobs}
A.E.~Jacobs, T.~Nikuni, H.~Shiba, J.Phys.Soc.Jpn. {\bf 62}, 4066 (1993); T.~Nikuni,
A.E.~Jacobs, Phys.Rev. B {\bf 57}, 5205 (1998).

\bibitem{weber}
H.B.~Weber, T.~Werner, J.~Wosnitza, H.v.L\"{o}hneysen, U.~Schotte, Phys.Rev. B {\bf
54}, 15924 (1996).

\bibitem{palme}
W.~Palme, F.~Mertens, O.~Born, B.~L\"{u}thi, U.~Schotte, Solid State Comm. {\bf 76},
873 (1990); W.~Palme, H.~Kriegelstein, G.~Gojkovic, B.~L\"{u}thi, Int.J.Mod.Phys.$_{
}$ B {\bf 7}, 1013 (1993).

\bibitem{ohta}
H.~Ohta, S.~Imagawa, M.~Motokawa, H.~Tanaka,\\
J.Phys.Soc.Jpn. {\bf 62}, 3011 (1993)
and Physica B {\bf 201}, 208 (1994).

\bibitem{tanaka}
H.~Tanaka, U.~Schotte, K.D.~Schotte, J.Phys.Soc.Jpn. {\bf 61}, 1344 (1992).

\bibitem{rastelli}
E.~Rastelli, A.~Tassi, Z.Phys.B {\bf 94}, 146 (1994).

\bibitem{wolf}
B.~Wolf {\it et al\/}., to be published.

\bibitem{date}
M. Date, M. Motokawa, A. Seki, S. Kuroda, K. Matsui, J.Phys.Soc.Jpn. {\bf 39}, 898
(1975);
S. Kuroda, M. Motokawa, M. Date, J.Phys.Soc.Jpn. {\bf 61}, 1036 (1992).

\bibitem{nojiri2}
H.~Nojiri, K.~Takahashi, T.~Fukuda, M.~Fujita, M.~Arai, M.~Motokawa, to be
published.

\end{references}
\end{document}